\documentclass[11pt,letter,oneside,twocolumn]{article}
\pdfoutput=1 
\usepackage[left=1.3cm,right=1.3cm,top=1cm,bottom=2cm]{geometry}
\usepackage[utf8]{inputenc}
\usepackage[english]{babel}
\usepackage{amsmath}
\usepackage{amsfonts}
\usepackage{amssymb}
\usepackage{mathtools}
\usepackage{placeins}
\usepackage{upgreek}
\usepackage[dvipsnames]{xcolor}
\usepackage{authblk} 
\usepackage{tabularx}
\newcolumntype{Y}{>{\centering\arraybackslash}X} 
\newcolumntype{Z}{>{\raggedleft\arraybackslash}X} 
\usepackage{booktabs}
\usepackage{abstract}
\usepackage[version=4]{mhchem}
\usepackage{graphicx} 
\usepackage{siunitx}
\DeclareSIUnit\arbitrary{a.u.}
\DeclareSIUnit\angstrom{\AA}
\DeclareSIUnit\rydberg{Ry}

\usepackage[backend=bibtex,style=numeric-comp, date=year, sorting=none, giveninits=true]{biblatex}
\AtEveryCitekey{\clearfield{title}}
\AtEveryBibitem{\clearfield{title}}
\renewbibmacro{in:}{%
  \ifentrytype{article}{}{\printtext{\bibstring{in}\intitlepunct}}} 

\usepackage{hyperref}
\newcommand{\abs}[1]{\left \lvert #1 \right \rvert}

\title{Excimer Formation in Zinc-phthalocyanine Revealed using Ultrafast Electron Diffraction}
\author[1]{Sebastian Hammer\thanks{mail to: sebastian.hammer@mail.mcgill.ca}}
\author[1]{Tristan L. Britt}
\author[1]{Laurenz Kremeyer}
\author[2]{Maximilian R\"{o}del}
\author[1]{David Cai}
\author[2,3]{Jens Pflaum}
\author[1]{Bradley Siwick\thanks{mail to: bradley.siwick@mcgill.ca}}

\affil[1]{Centre for the Physics of Materials, Departments of Physics and Chemistry, McGill University, Montreal, Canada}
\affil[2]{Experimental Physics 6, Julius-Maximilians University W\"{u}rzburg, W\"{u}rzburg, Germany}
\affil[3]{CAE Bayern, W\"{u}rzburg, Germany}
\date{}
\begin{document}

\twocolumn[
  \begin{@twocolumnfalse}
\maketitle
	\begin{abstract}
        The formation of excited dimer states, so called excimers, is an important phenomenon in many organic molecular semiconductors. In contrast to Frenkel exciton-polaron excited states, an excimer is long-lived and energetically low-lying due to stabilization resulting from a substantial reorganization of the inter-molecular geometry. In this letter, we show that ultrafast electron diffraction can follow the dynamics of solid-state excimer formation in polycrystalline thin films of a molecular semiconductor, revealing both the key reaction modes and the eventual structure of the emitting state. We study the prototypical organic semiconductor zinc-phthalocyanine (ZnPc) in its crystallographic $\upalpha$-phase as a model excimeric system. We show, that the excimer forms in a two-step process starting with a fast dimerization ($\lesssim \SI{0.4}{\ps}$) followed by a subsequent slow shear-twist motion ($\qty{14}{\ps}$) leading to an alignment of the $\uppi$-systems of the involved monomers. Furthermore, we show that while the same excimer geomtry is present in partially fluorinated derivatives of ZnPc, the formation kinematics slow down with increasing level of fluorination. 
	\end{abstract}
\end{@twocolumnfalse}
]
\saythanks
Optical excitation of molecular semiconductors typically yields Frenkel excitons, where both the hole and the electron are located on a single molecular unit due to the large Coulomb interaction energy that results from the low dielectric constants of these materials \cite{Schwoerer2007,Davydov1948}.
Local coupling to molecular vibrations extends the Frenkel picture to an exciton-polaron description \cite{Spano_2010,Hestand_2018} which captures many additional experimental observations such as superradiance in J-aggregates \cite{Spano_2011,M_ller_2013}.
Beyond vibrational coupling of excited monomers, there are other excited states of molecular semiconductors that are bi-molecular in nature and involve a significant reorganization of the local molecular arrangement.  
Excimers (from excited dimer) are perhaps the foremost example of such excitations. 
In these cases, the mixing of charge-transfer (CT) and Frenkel excitons on adjacent molecules can lead to a new inter-molecular geometry within the crystal, resulting in a low-lying excited state on a dimer unit rather than a monomer as expected in the Frenkel picture. 
Excimer states tend to be associated with  a broad emission spectrum \cite{Engels_2017,Bialas_2022} making them an active research subject in broadband emission organic light emitting diodes \cite{Kalinowski_2009,Reineke_2013, Vollbrecht_2018, Hammer_2019}. 
While many theoretical studies have investigated the geometric changes involved in the formation of an excimer state \cite{Fink_2008,Casanova_2015,Hoche_2017}, experimental investigations have concentrated on optical and diffusion properties \cite{Pensack_2018,Rehhagen_2023} or inferred changes in the inter-molecular geometry by tracking vibrational modes during excimer formation \cite{Hong_2022} rather than directly probing the changes in structure. \par
\begin{figure*}[ht]
  \includegraphics{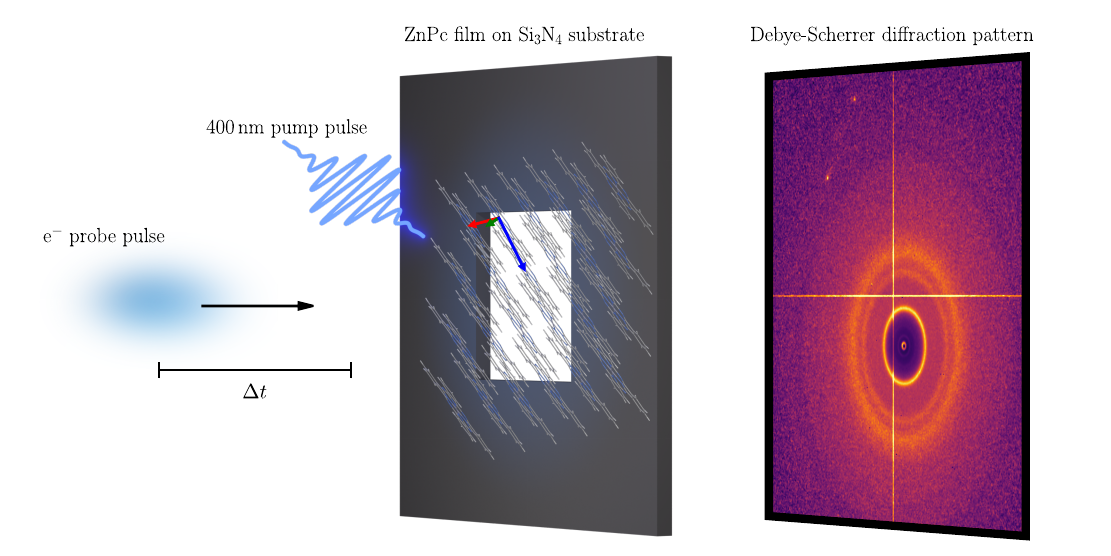}
  \caption{Schematic depiction of the pump-probe electron diffraction experiment. A ZnPc thin film on a Si$_3$N$_4$ electron diffraction window is excited with a fs pump pulse of \qty{400}{\nm} center wavelength. After a delay time $\Delta t$, the momentary crystal structure is probed by a fs electron pulse. Due to the polycrystalline nature of the thin film, the resulting diffraction pattern presents as Debye-Scherrer rings. The shown intensity map was weighted by $q^2$. The crystallographic \textbf{a},\textbf{b} and \textbf{c}-axes are shown for a representative unit cell as a red, green and blue arrow.}
  \label{fig:1}
\end{figure*}
Here, we use ultrafast electron diffraction (UED) to directly follow the structural dynamics involved in solid-state excimer formation, extending previous UED work on structural dynamics in molecular single crystals\cite{Gao_2013,Ishikawa_2015,Liu_2017,Smit_2019,Seiler_2021} to technologically relevant polycrystalline thin films.  
Specifically, we determine the geometric changes and time-constants associated with excimer formation in the prototypical molecular semiconductor zinc-phthalocyanine (ZnPc) in its thin film crystallographic $\upalpha$-phase for various degrees of fluorination of its periphery, thereby intentionally modifying the inter-molecular potential by the variation of molecular mass and van der Waals radii. 
ZnPc is well studied as a donor in organic photovoltaic devices due to its durability, favorable absorption spectrum and the tunability of the energetic position of its molecular frontier orbitals by fluorination without strongly changing the optical gap \cite{Bruder_2010,Meiss_2011,Tietze_2013,Brendel_2015,Schwarze_2016}.
While the first excited state is predicted to have a strong CT admixture \cite{Feng_2022}, a fact that is corroborated by luminescence studies on these films which show a strong excimer signature \cite{Hammer_2023}, the structural changes involved in excimer formation have been elusive to experimental and theoretical studies so far.
With UED we show that these dynamics primarily involve a two-step process of fast dimerization followed by a slower shear-twist of the dimer embedded in the crystalline environment.  Our studies also reveal significant conformational flexibility on the excited state potential energy surface that provides a possible dynamical basis for the broad emission. While the underlying geometric distortion is unaffected by the level of fluorination, the formation kinetics slow down significantly with increasing fluorination.\par
Thin films of ZnPc, four-fold (F$_4$ZnPc) and eight-fold (F$_8$ZnPc) fluorinated ZnPc of \qty{50}{\nm} thickness were deposited on silicon nitride Si$_3$N$_4$ TEM windows (area $\qty{250}{\um}\times \qty{250}{\um}$, thickness \qty{30}{\nm}) via organic molecular beam deposition under high vacuum (\qty{1e-7}{\milli\bar}) from two-fold gradient-sublimed purified source material. 
Polycrystalline F$_n$ZnPc films grows in a metastable crystallographic $\upalpha$-phase \cite{Brendel_2015, Hausch_2023}. 
Due to the in-plane rotational domain texture of the polycrstalline thin films, the diffraction pattern consists of Debye-Scherrer rings as shown in the diffraction pattern on the right hand side of figure~\ref{fig:1}. In the kinematic diffraction picture, the integrated intensity for an \textit{hkl}-reflection of a true (untextured) powder sample is given by \cite{Zuo_2019}
\begin{equation}
  I_{hkl}(q_{hkl},t) = I_0 m_{hkl}\frac{\lambda^2 d_{hkl} V_\text{sample}}{2V_\text{c}^2} \abs{F_{hkl}(q_{hkl},t)}^2,
  \label{eq:1}
\end{equation}
with $q_{hkl}$ being the scattering vector satisfying the Laue-condition for the $hkl$-reflection with multiplicity $m_{hkl}$.
Furthermore, $I_0$ is the incident intensity, $\lambda$ is the de~Broglie wavelength of the electrons, $d_{hkl}$ is the lattice distance of the corresponding $hkl$-reflection and $V_\text{c}$ and $V_\text{sample}$ are the unit cell and sample volume, respectively. 
The scattering intensity is determined by the structure factor
\begin{equation}
  F_{hkl}(q_{hkl},t) = \sum_s f_s(q_{hkl}) e^{-M_s} e^{-2\pi \i \mathbf{\overline{q}}_{hkl} \cdot \mathbf{x}_s(t)}.
  \label{eq:2}
\end{equation}
The sum counts each atom,~$s$, in the unit cell and $f_s(q_{hkl})$ is the respective atomic form factor for electron scattering.
The phase factor~$e^{-2\pi \i \mathbf{\overline{q}}_{hkl} \cdot \mathbf{x}_s(t)}$ is dependent on the fractional atom position~$\mathbf{x}_s(t)$ in the unit cell at time~$t$ with $\mathbf{\overline{q}}_{hkl}=(h,k,l)$. 
$M_s$ decribes the Debye-Waller effect caused by incoherent displacements of atoms due to their thermal movement around the equilibrium position~$\mathbf{x}_s(t)$. \par
Figure~\ref{fig:1} shows a schematic of the pump-probe UED experiment, performed in transmission geometry \cite{Chase2016,Waldecker2017,Stern2018}. 
An optical pump-pulse (\qty{400}{\nm} center wavelength, fluence \qty{0.5}{\milli\joule\per\cm\squared}) excites the molecules within the film into the B-band (c.f. SI figure~1). This band comprises excitations to higher singlet states $S_0 \rightarrow S_{n\geq2}$ \cite{Wallace_2017}.
From there, according to Kasha's rule \cite{Kasha1950, Siebrand_1967}, the excitation quickly relaxes to the longer lived $S_1$ state from which the excimer is formed. 
These excitation conditions lead to an eventual excimer density of about one in a thousand molecules for ZnPc and one in a hundred for F$_4$ZnPc and F$_8$ZnPc (for details see SI section~1).
The optical pump-pulse is followed by a radio frequency compressed electron probe ($\approx 10^{4} \text{ e}^-$ per pulse, \qty{300}{\fs}) using an instrument desribed in detail elsewhere \cite{Chatelain_2012,Otto_2017}. The electron diffraction pattern is measured as a function of pump-probe time delay $\Delta t$, $I_{hkl}(q, \Delta t)$, from which the changes in diffraction intensity resulting from optical excitation can be determined. We use the measured intensities to infer the changes in structure involved in the excimer formation in a manner similar to that used to interrogate structural dynamics following optical excitation in nanocrystalline lead halide perovskite materials \cite{Wu_2017, Seiler_2023, Yazdani_2023}, vanadium dioxide \cite{Morrison_2014, Otto_2018} and other materials \cite{hauf2023}.
As previously shown \cite{Britt_2022}, Si$_3$N$_4$ shows no signals of structural dynamics upon excitation with \qty{400}{\nm}, which is why we can attribute the observed changes in diffraction intensity exclusively to the molecular films. \par
\begin{figure}[btph]
  \includegraphics{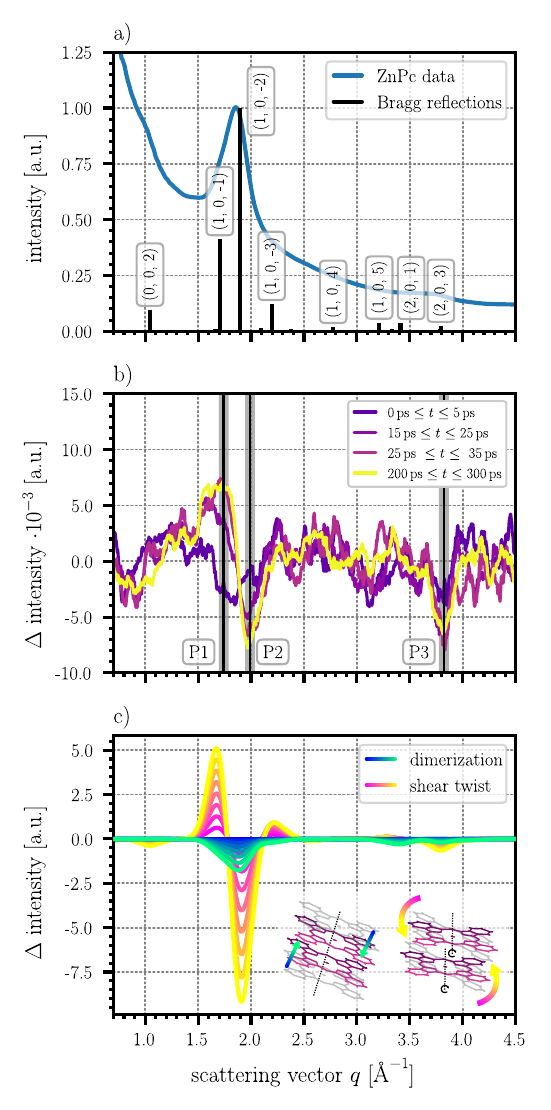}
  \caption{a) Angular integration of ZnPc diffraction pattern (blue). 
  Bar plot represents major Bragg reflections contributing to the observed diffraction data. 
  b) Relative transient diffraction data for different time intervals. 
  Main features are marked by black lines labeled P1 to P3. 
  Grey area marks averaged $q$-range for time traces in figure~\ref{fig:2} c). 
  c) Simulated transient change for dimerization and shear-twist. Geometric relaxation shown as insets and in figures~\ref{fig:4} a) and b).
  }
  \label{fig:2}
\end{figure}
\begin{figure*}[bth]
  \includegraphics{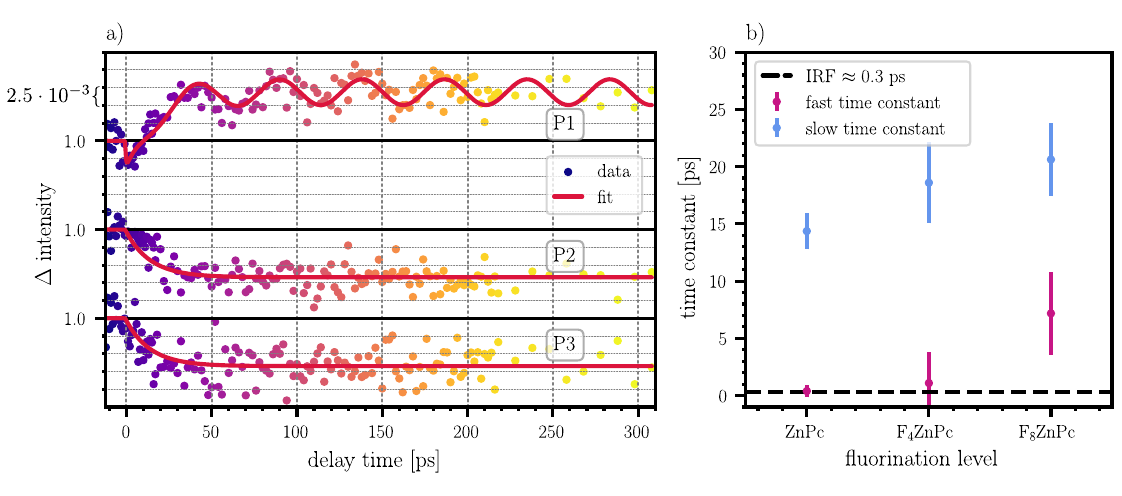}
  \caption{
  a) Time traces of the three main features P1, P2 and P3 (top to bottom). 
  The solid line indicates the kinetic fit based on changes in the diffraction signal induced by the dimerization and shear-twist motion indicated by the structural model.
  b) Time constants of the excimer formation process as a function of fluorination. 
  The black dashed line marks the instrument response function of width of \qty{300}{\fs}.
  The fast initial time constant associated with the dimerization of adjacent molecules, while the slow time constant is covering the relaxation process governed by a shearing motion towards the final geometry.}
  \label{fig:3}
\end{figure*}
Figure~\ref{fig:2} a) shows the angular integrated intensity of a ZnPc diffraction pattern using the \texttt{pyFAI} software package \cite{Kieffer_2020}. 
The observed powder diffraction pattern shows a weak peak at \qty{1.0}{\per\angstrom} and a strong peak at \qty{1.9}{\per\angstrom} in agreement to the in-plane crystal structure of the ZnPc $\upalpha$-phase \cite{Hausch_2023}. 
For comparison, the black bars in figure~\ref{fig:2} a) indicate the position and intensity of the strongest Bragg reflections that contribute to the in-plane diffraction signal at room tmperature, derived from the published copper-phthalocyanine unit cell \cite{Erk_2004} and a full quantum chemical unit cell relaxation using \texttt{QuantumESPRESSO} \cite{Giannozzi_2009,Giannozzi_2017} (see SI section 2 for details). 
We now turn our attention to the relative change in diffraction intensity $\Delta I(q, t) = (I(q, t)-I_0(q))/I_0(q)$, normalized with respect to the equilibrium intensity $I_0(q)$ before photo-excitation at $t \le \qty{0}{\ps}$. 
These differential diffraction signals, shown in figure~\ref{fig:2} b), contain fingerprint signatures of the structural changes involved in ZnPc excimer formation.
At long times $t \geq \qty{200}{\ps}$ we find a stable pattern that is composed of three prominent features marked by black lines labeled P1 to P3, with intensity increasing at about  $q=\qty{1.7}{\per\angstrom}$ (P1) and intensity decreasing at $q=\qty{2.0}{\per\angstrom}$ (P2) and $q=\qty{3.8}{\per\angstrom}$ (P3). The increase at P1 is a clear indication of a change in structure, since the Debye-Waller contribution that describes random thermal motion in equation \eqref{eq:2} only leads to a decrease in intensity. 
We emphasize that due the low excitation levels used in our study, the largest recorded signal changes are well below \qty{1}{\percent}.\par
Next, we turn our attention to the kinetics of the observed signal changes.
In figure~\ref{fig:3}~a) the time traces of the features P1 to P3 are presented.
At P1, we find a fast signal decrease ($\approx \qty{1}{\ps}$) followed by a slower signal increase ($\approx \qty{15}{\ps}$) which is superimposed by a slow oscillation, whose origin we will discuss later in the manuscript.
For P2 and P3 the diffraction intensity is dominated by a slow decrease, stabilizing after \qty{50}{\ps} and showing no further changes within in the measured time range of \qty{300}{\ps}. 
Interestingly, we find that the qualitative features of the observed intensity changes for F$_4$ZnPc and F$_8$ZnPc (see SI section 6) are identical to those for ZnPc, strongly suggesting that i) the same reaction modes (or structural distortions) stabilize the final excimer state regardless of fluorination level and that ii) the structure of the emissive state is essentially identical in all three cases.\par
\begin{figure*}[tbh]
  \includegraphics{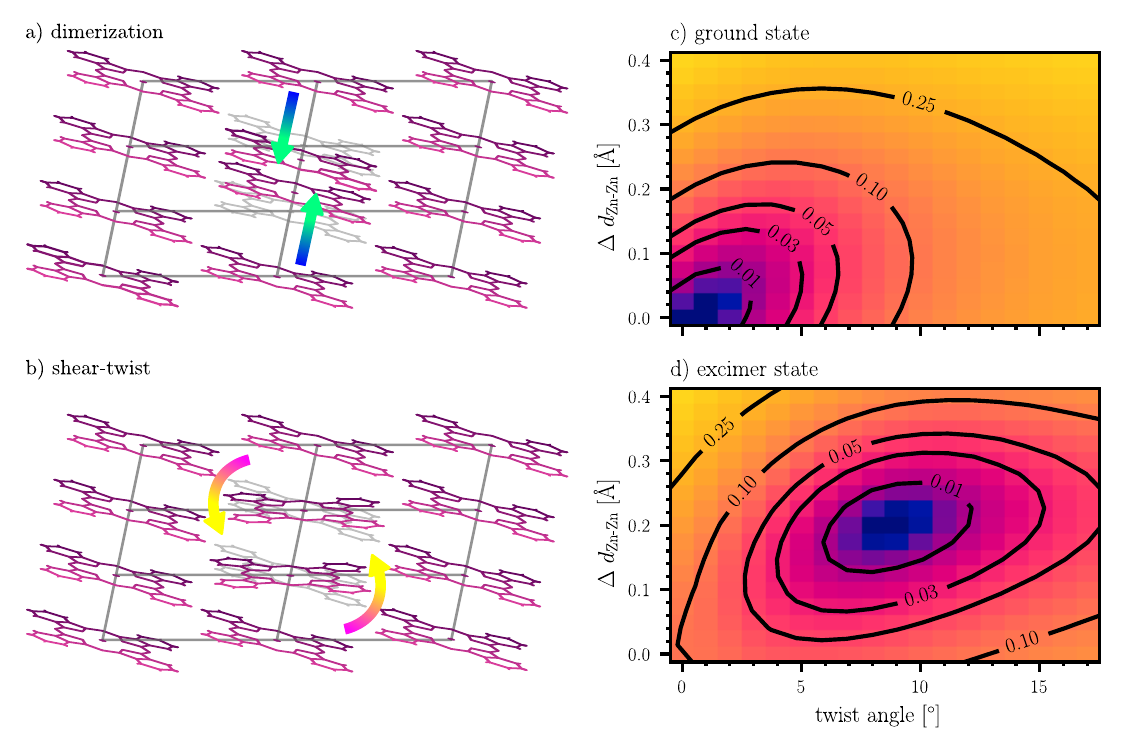}
  \caption{
  Structural model of a) dimerization of two adjacent molecules and b) shear-twist of two adjacent molecules in a crystalline environment. 
  Potential energy surfaces of c) the ground and d) the excimer state of the isolated dimer as a function of decreasing inter-molecular dimer distance $\Delta d_\text{Zn-Zn}$ and twist angle. Contour line values in eV relative to the respective minimum. The ground state has its minimum at the equilibrium geometry of the crystal. The excimer state has its minimum at a twist angle of \qty{9}{\degree} and a decrease of dimer distance of \qty{0.2}{\angstrom}.}
  \label{fig:4}
\end{figure*}

We model the impact of a broad set of possible molecular displacements on the powder diffraction intensity from $\upalpha$-ZnPc crystallites in an attempt to describe observations presented in figure~\ref{fig:2}. All of the displacements modeled have been suspected to play a role in the excimer formation of planar molecules.
While dimerization, i.e. the convergence of two adjacent molecules along the direction of the $\uppi$-systems, is widely accepted to be a part of the final excited-state geometry, a more significant change in the alignment of the $\uppi$-systems resulting from lateral shearing or rotations are also potentially relevant motions during the excimer formation \cite{Birks1968,Warshel_1974,Pensack_2018,Fink_2008,Casanova_2015,Hoche_2017}.
Hence, in addition to a dimerization coordinate, we considered a set of in-plane and out-of-plane shearing motions and rotations around the central normal axis of the molecule to find the minimal set of motions that can provide a quantitative match to the data within the limits of the singal-to-noise ratio.
This analysis was performed in an unbiased manner, simulating the changes in diffraction intensity across the entire range of observed scattering vectors as a function of the displacement distance (from the equilibrium geometry) for each distinct displacement described (see SI, section 3). 
Comparing the predicted impacts with observed changes in relative intensity, it is clear by visual inspection that most of these proposed motions are in profound qualitative disagreement with the observations.
Out of all the investigated motions, only one combination provides qualitative agreement with the observed changes in diffraction intensity:
(i) a dimerization along the crystallographic \textbf{a}-direction (short in-plane axis) that increases the overlap between the $\uppi$-systems of the adjacent monomers (c.f. figure~\ref{fig:4} a)) and (ii) a shear-twist motion around an out-of-plane symmetry axis pinned at the central zinc atom which leads to improved alignment of the $\uppi$-systems of the adjacent monomers (c.f. figure~\ref{fig:4}~b)).\par
In figure~\ref{fig:2}~c) we show the changes in the diffraction intensity associated with dimerization and shear-twist motion (indicated in the inset). 
Here, decreases in inter-molecular distance, i.e. the zinc-zinc distance between adjacent monomers, is shown from blue (\qty{0}{\angstrom}) to green (\qty{0.2}{\angstrom}).  Increases in twist angle are shown by the curves from pink (\qty{0}{\degree}) to yellow (\qty{4}{\degree}). Combined, these are the only displacement coordinates that can explain the fast reduction in intensity observed in the P1 region, followed by the the slower positive (P1) and negative (P2, P3) going features observed.
The intensity changes observed at long times are clearly dominated by a shear-twist displacement of the dimer, for which we find a large signal increase at $q=\qty{1.7}{\per\angstrom}$ and a signal decrease at $q=\qty{1.9}{\per\angstrom}$ matching both the P1 and P2 features at late delay times. For the P3 feature we find a matching signal decrease around $q=\qty{3.8}{\per\angstrom}$. 
The changes observed at long delay times, over the entire range of scattering vectors observed, are reproduced with surprising accuracy by this shear-twist motion alone. Thus, we confidently conclude that the dominant structural reorganization involved in excimer formation in $\upalpha$-crystalline ZnPc and its flurorinated derivatives is the shear-twist movement shown in figure~\ref{fig:4}~b).
However, turning our attention to the early time scales for P1 we see that the fast intensity decrease cannot be explained by the shear-twist motion, since only intensity increases are found in this $q$-region for this structural distortion. 
Displacements along the inter-molecular dimerization direction, however, produce negative going intensity changes with the largest contribution around $q=\qty{1.9}{\per\angstrom}$ as observed on the picosecond timescale in the P1-P2 region of the data (figure~\ref{fig:2}~b) purple, $\qty{0}{\ps} \leq t \leq \qty{5}{\ps}$). 
Thus, a two-step excimer formation process that involves fast dimerization followed by subsequent shear-twist motion to stabilize the dimer in the crystalline environment provides a minimal quantitative description of our UED data.\par 
From this structural model we can now extract the kinetics of the excimer formation from the time traces in figure~\ref{fig:3}~a). 
We fit a global model, using a bi-exponential decay for P1 and P2 to capture the dimerization and shear-twist motion and a mono-exponential decay for P3 to account for the shearing-twist present at this $q$-value.
The details of this procedure are given in the SI section 4. 
We find that the dimerization occurs on a fast time scale below \qty{0.4(5)}{\ps} while the shearing motion is much slower with a time constant of \qty{14.4(16)}{\ps}.
While the time constant for the dimerization is fast, it is not uncommon for such a fast geometrical relaxation to take place \cite{Schubert_2013,Hoche_2017,Hong_2022}.
Optical studies on related materials have suggested that the excimer formation is governed by a two-step process where the initial formation yields a ``hot'' excimeric state which relaxes to the equilibrium excimer geometry on the time scale of tens of picoseconds \cite{Pensack_2018,Hong_2022}. 
We likewise conclude that the excimer formation in ZnPc in its crystallographic $\upalpha$-phase is governed by a two-step process. The initial optical excitation is followed by a fast dimerization. 
Subsequently, this excited dimer state relaxes to the excimer geometry by a shear-twist motion. The final geometry then persists beyond \qty{300}{\ps}.
Recently, Bialas and Spano hypothesized \cite{Bialas_2022}, that a change in overlap of the $\uppi$-orbitals of adjacent monomers along a 'slow' inter-molecular coordinate can act as a non-local coupling mechanism between the Frenkel and CT state and stabilize the excited dimer state in a new geometry with a significant reduction in energy. 
The emission takes place from this state, the excimer state, yielding a broad spectrum due to vibronic coupling to low-energy inter-molecular vibrational modes. 
The shear-twist motion we observe provides such a change in the overlap of the $\uppi$-orbitals of the adjacent ZnPc monomers. We therefore suggest, that this shear-twist is the dominant inter-molecular motion that stabilizes the emissive excimer state in its distorted geometry.
\par
With this kinetic model at hand for ZnPc, we analyze the excimer formation kinetics as a function of fluorination. 
The transient diffraction data for F$_4\, $- and F$_8$ZnPc, including the time traces, is shown in the SI section 5.
Looking at the time traces of the key features P1 to P3, we find similar behavior indicating a two-step process with the same origins as in ZnPc, however with different time constants. 
Hence, we use the same kinetic model as for ZnPc to extract the time constants for the dimerization and shear-twist distortion. 
While for the fast initial step, we find that the time constants of ZnPc and F$_4$ZnPc are close to our time resolution, we see a significant increase of an order of magnitude for F$_8$ZnPc. 
The slower time constant governing the relaxation towards the final excimer geometry along the shear-twist degree of freedom shows a clear trend towards larger values with increasing level of fluorination. 
We attribute this retardation of the excimer formation to the larger van der Waals radius of the fluorine atoms and the enhanced molecular weight modifying the inter-molecular interaction which leads to an enhanced steric hindrance as well as a slowing down of the promoting phonon modes of the involved molecular movements.\par
Having shown that the structural deformation accompanying the formation of excimers appears to be dominated by dimerization and shear-twist, we investigate the impact of this structural reorganization on the free energy of the crystal after photoexcitation of its molecular entities using density functional theory methods.
We perform ground and excited state self-consistent field calculations using \texttt{QuantumESPRESSO}\cite{Giannozzi_2009,Giannozzi_2017} for an isolated ZnPc dimer in different frozen geometries.
We follow the method suggested by Marini and Calandra \cite{Marini_2021} using two separate chemical potentials to parameterize the electrons and holes in the excited state, with one electron in the conduction band and one hole in the valence band manifold. 
The intra-molecular geometry of the monomer units was optimized to encourage shared excitation across the dimer entity as outlined in e.g. \cite{Craciunescu_2022,Craciunescu_2023}.
The dimer is described as the dimer geometry found in the crystal along the crystallographic $\mathbf{a}$-axis.
The dimerization is parameterized as the decrease in zinc-zinc distance $\Delta d_\text{Zn-Zn}$ and the shear-twist is expressed by the twist angle of both molecules around the out-of-plane axis (c.f. figure~\ref{fig:4}~a) and~b), respectively.
The full computational details are given in the SI section 5. 
Figure~\ref{fig:4}~c) shows the ground state potential energy surface (PES) as a function of these two parameters. 
We find that the ground state crystal geometry is also the energetic minimum in the explored geometric space.
For the excited state, the PES is presented in figure~\ref{fig:4}~d).
The global minimum of the excited PES occurs for a decrease in dimer distance of~$\approx \qty{0.2}{\angstrom}$ and a twist angle of~$\approx\qty{9}{\degree}$.
While these exact values can differ in the crystal due to the steric hindrance by neighboring molecules, the calculation on the isolated dimer reproduces the observed relaxed excimer configuration from the time-resolved diffraction data, further supporting the suggested two-step model.
From a purely energetic perspective, the calculated excited state PES does explain the observed final geometry very well.
However, it fails to explain the large difference in the two time constants attributed to the movement along the two generalized coordinates.
Moving in free energy along a pure dimerization from the ground state minimum, i.e. the ground state crystal structure, increases the total energy of the system, while the shear-twist motion decreases the total energy. 
This suggests that the relaxation would follow a path where initial movement should happen along the shear-twist coordinate, followed only later by a dimerization towards the global energy minimum. 
The performed calculations however are restricted in that sense, that they assume a fully relaxed electronic system which is not the case in the first few hundred femtoseconds. 
Studies on comparable molecular systems \cite{Schubert_2013,Schubert_2014} suggest that the initial fast distortion of the dimer geometry goes hand-in-hand with a mixing of the Frenkel and CT states preparing the initial hot excimer state from which the subsequent structural relaxation takes place.\par
Finally, we want to return to the coherent signal changes we observe in the time traces of P1. Oscillations in the intensity of a Bragg reflection can be caused by the excitations of coherent phonons as they periodically modify the atomic equilibrium positions $\mathbf{x}_s$, which determine the structure factor given by equation \eqref{eq:2}. 
For a system that changes its equilibrium geometry in the excited state, the relaxation towards the new geometry can be accompanied by emerging phonon modes around the new PES minimum, similar to the mechanism of displacive excitation of coherent phonons \cite{Zeiger_1992}. 
As discussed above, we find a new PES minimum displaced from the ground state geometry along the dimerization and shear-twist coordinate. 
However, coherent movements along these coordinates should not only be observed at P1 but also at P2. 
An alternative explanation is the population of coherent acoustic phonons associated with out-of-plane heat transport which then modulate the local atomic geometry, a common phenomena in UED experiments\cite{Chatelain_2014,Zhang_2023,Ungeheuer_2024}. 
From the oscillation frequency, we can estimate the speed of sound in the out-of-plane direction by assuming the lowest quantization of the out-of-plane accoustic phonon mode, i.e. the film thickness is half the wavelength (see SI section 4 for details). 
This yields \qty{2.07(11)}{\km\per\s} which is within a reasonable range compared to other organic small molecule crystals \cite{Schwoerer2007,Roua_1999,Kamencek_2022}.
We hence suggest, that this is the likely explanation for the observed oscillation of the diffraction signal, but want to point out that we cannot exclude any displacive process along a minor distortion coordinate we did not consider.\par
In this letter we have shown how UED can be used to directly determine the structural changes associated with excimer formation in crystalline molecular aggregates in a manner equivalent to making an ultra-fast molecular movie of this solid-state bi-molecular reaction. 
Previously, all dynamical processes facilitating the formation of the technologically relevant excimer state were indirectly inferred from spectroscopic measurements. 
Here we demonstrate, for the first time, how to directly access the nuclear part of the excimer formation process on its intrinsic time-scales.
Structural models that emerge from an analysis of UED data can be used as a basis to guide and inform quantum chemical calculations from which additional details can be gleaned.  
The low excitation density enabled us to study a minimally perturbed system in which subtle changes in the geometry are not hidden by incoherent Debye-Waller dynamics. 
We uncovered that the excimer formation in the prototypical molecular semiconductor ZnPc and its partially fluorinated derivatives is governed by a two-step process: (i) a fast dimerization of adjacent monomers followed by (ii) a subsequent shear-twist motion which aligns the $\uppi$-systems of the participating ZnPc units. 
This model is supported by quantum chemical calculations of the ground and excited state PESs over a large parameter space.
These structural dynamics persist upon four- and eight-fold fluorination of the molecular ZnPc periphery. 
However, the geometric relaxation facilitating the excimer formation is damped by the presence of the large fluorine atoms. 
The ability to access and understand the transient structural and optical processes in these materials is of utmost importance in paving the way to the development of highly efficient organic opto-electronics.
\FloatBarrier
\section*{Acknowledgements}
S.H. gratefully acknowledges funding from the German Research Foundation within project~490894053. 
This work was supported by the NSERC Discovery Grants Program under Grant No. RGPIN-2019-06001 (B.J.S.). 
This work was also supported by the Fonds de Recherche du Québec-Nature et Technologies (FRQNT), the Canada Foundation for Innovation (CFI), and the National Research Council of Canada (NRC) Collaborative R\&D program (Quantum Sensors).
L.K. and T.L.B. gratefully acknowledge the support from an FRQNT Merit fellowship.
J.P. and M.R. acknowledge financial support by the Bavarian State Ministry for Science and the Arts within the collaborative research network “Solar Technologies go Hybrid” (SolTech).
We thank Prof. Bernd Engels for the helpful discussion.
\printbibliography
\end{document}


\maketitle
\tableofcontents

\section{Exciton density}
\label{sec:1}
\begin{figure}[tb]
  \includegraphics{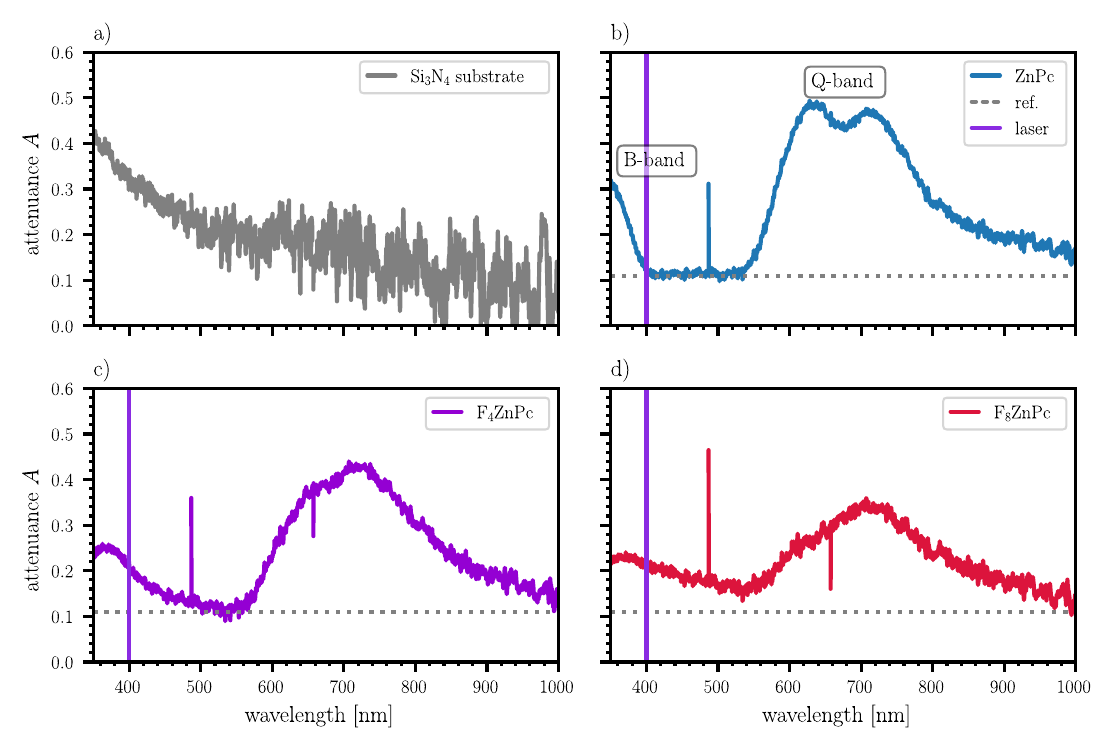}
  \caption{Attenuance spectra of a) reference S$_3$N$_4$ substrate, b) ZnPc thin film, c) F$_4$ZnPc thin film and d) F$_8$ZnPc thin film. The excitation laser wavelength is shown as a purple line. The baseline used as a zero reference for the attenuance ratio is shown in grey.}
  \label{fig:S1}
\end{figure}

To estimate the upper limit of exciton density we need the fraction of photons that is absorbed by the molecular thin film. We performed transmission spectrum measurements on the thin film samples of ZnPc, F$_4$ZnPc and F$_8$ZnPc each of \qty{50}{nm} thickness on Si$_3$N$_4$ TEM substrates with a \qty{30}{nm} thick Si$_3$N$_4$ square $\qty{250}{\micro\meter}\times\qty{250}{\micro\meter}$ membrane. 
This gives us the attenuance as the complementary quantity to the transmittance $T$ 
\begin{equation}
  A = 1 - T 
  \label{eq:S1}
\end{equation}
where we assume that the emittance $E\approx0$ is negligible.  
The transmittance of a film of thickness $d$ is the ratio of the transmitted intensity $I(d)$ and the initial intensity $I_0$ and is described by the Lambert-Beer law \cite{Parson2007} 
\begin{equation}
  T=I(d)/I_0=\exp\left(-d\alpha\right)
  \label{eq:S2}
\end{equation}
where $\alpha$ is the extinction coefficient in units of inverse length. 
We measured the transmission spectra using a home-built UV-NIR transmission spectrophotometer. It utilizes a deuterium tungsten-halogen lamp (Ocean Optics DH-2000) to provide broadband illumination. 
The output light is initially fiber-coupled onto two parabolic silver mirrors, which then collimate and focus the beam down to a spot size at the sample position of less than \qty{50}{\micro\meter}. 
Subsequently, a symmetric setup with two parabolic mirrors focuses the beam into the fiber-end of an OceanOptics USB2000 spectrometer (\qty{200}{\nm} to \qty{1100}{\nm}).
We make the assumption, that the fraction of absorbed photons is given by the attenuance $A$, neglecting all reflective contributions from all interfaces for this sample geometry.
Under these assumptions we can determine the transmittance of the neat molecular film $T_M$ as 
\begin{equation}
  T_\text{M} = \frac{T_\text{S}}{T_\text{SiN}}.
  \label{eq:S3}
\end{equation}
Here $T_\text{S}$ is the transmittance of the stacked film system and $T_\text{SiN}$ the transmittance of the neat  Si$_3$N$_4$ substrate.
Figure \ref{fig:S1} shows the attenuance calculated according to equation \eqref{eq:S1} for the Si$_3$N$_4$ substrate a), the ZnPc b),  F$_4$ZnPc c) and F$_8$ZnPc d) films. 
For the molecular films, we see that the attenuance does not drop to zero, even for the region between the Q- (around \qty{700}{\nm}) and the B-band (above $\approx \qty{450}{\nm}$) where there are no molecular transitions \cite{Theisen_2015}. 
We attribute this to the occurrence of reflection effects at the interfaces omitted in our considerations, as the band gap of Si$_3$N$_4$ is large and hence all the attenuance found for the substrate (c.f. figure \ref{fig:S1} a) can be attributed to reflections on the front or back surface. 
To estimate the actual fraction of absorbed photons we hence will use the plateau between \qty{420}{\nm} and \qty{520}{\nm} as the reference value $A=0$. 
With that we find for the attenuance of the molecular films $A_\text{ZnPc} \approx 0.01$, $A_\text{F$_4$ZnPc} \approx 0.1$ and $A_\text{F$_8$ZnPc} \approx 0.1$. \par
Following the procedure in Seiler \textit{et al.} \cite{Seiler_2021} we can now estimate the fraction of excited molecules per laser pulse. Using the photon energy 
\begin{equation*}
  E_\text{ph} = \frac{hc}{\lambda} = \frac{\qty{2e-25}{\joule\meter}}{\qty{400e-9}{\m}} = \qty{5e-19}{\joule}
\end{equation*}
we can calculate the number of incident photons from the incident flux $F=\qty{0.5e-3}{\joule\per\cm\squared}$ as
\begin{equation*}
  N_\text{ph} = \frac{F}{ E_\text{ph}} = \qty{1e15}{\per\cm\squared}.
\end{equation*}
The number of absorbed photons per unit volume can be calculated using the attenuance $A$ and the film thickness $d$. For ZnPc, this yields
\begin{equation*}
  n_\text{ZnPc} = \frac{A_\text{ZnPc} \cdot N_\text{ph}}{d} = \frac{0.01\times \qty{1e15}{\per\cm\squared}}{\qty{50e-7}{\cm}}\approx \qty{2e18}{\per\cm\cubed}.
\end{equation*}
With the ZnPc unit cell volume of roughly $V_\text{uc,ZnPc}=\qty{600}{\cubic\angstrom}$ and one molecule per unit cell the fraction of excited molecules in the film is
\begin{equation*}
  \eta_\text{ZnPc} =  n_\text{ZnPc}\cdot V_\text{uc,ZnPc} = \qty{2e18}{\per\cm\cubed} \times \qty{600e-24}{\cubic\cm} \approx 0.0012.
\end{equation*}
Hence, every thousandth molecule is excited at the incident fluence chosen here. For the fluorinated counterparts, the absorption is roughly a factor of ten larger than for ZnPc, yileding that every hundredth molecule is excited in these films. 

\FloatBarrier
\section{ZnPc ground state geometry from density functional theory}
As no crystal structure for the ZnPc $\upalpha$-polymorph is available in literature, we performed a structural optimization within a density functional theory framework using \texttt{QuantumEspresso} \cite{Giannozzi_2009,Giannozzi_2017} starting from the morphological similar Copper-phthalocyanine (CuPc) crystal structure\cite{Erk_2004,Erk_2004_cif}.

Calculations used projector augmented wave (PAW) pseudopotentials with nonlinear core corrections with the Perdew-Burke-Ernzerhof functional \cite{Perdew1996} and the kinetic and wave function energy cutoffs were \qty{600}{\rydberg} and \qty{60}{\rydberg}, respectively. A semi-emperical
three-body correction was applied following the method of Grimme \cite{Grimme_2010,Grimme2011} to account for van der Waals dispersion, with Becke-Johnson damping to ensure the convergence of the correction at short distances \cite{Becke2005}. Electronic structure calculations were done on an $4\times 4\times 4$ Monkhorst-Pack $\mathbf{k}$-grid.
To obtain the ground state structure, the geometric relaxation of the unit cell was performed in two steps. First, a full cell relaxation keeping the initial symmetry  at P$\overline{1}$ using Parrinello-Rahmen molecular dynamics \cite{Parrinello1981} was carried out. After that, the atomic positions within the new cell were relaxed using damped Verlet-integration dynamics. Only minor differences to the CuPc crystal structure were found (c.f. table \ref{tab:1}).
\begin{table}[htb]
\centering
\begin{tabular}{l | r | r }
  cell parameter& CuPc (origin) &ZnPc (optimized) \\ \hline \hline
  a & \qty{3.8052}{\angstrom} & \qty{3.7857}{\angstrom}\\ \hline
  b & \qty{12.9590}{\angstrom}& \qty{12.9785}{\angstrom}\\ \hline
  c & \qty{12.0430}{\angstrom}& \qty{12.0525}{\angstrom}\\ \hline
  $\upalpha$ &\qty{90.6400}{\degree} &\qty{90.5357}{\degree} \\ \hline
  $\upbeta$ &\qty{95.2600}{\degree} & \qty{94.5759}{\degree}\\ \hline
  $\upgamma$&\qty{90.7200}{\degree} & \qty{90.7964}{\degree}
\end{tabular}
\caption{Comparison of original and DFT optimized crystal structures.}
\label{tab:1}
\end{table}

\section{Systematic study of relaxation geometries}
To understand the observed transient diffraction pattern shown in figure 1 b) in the main manuscript, we considered several possible relaxation geometries apart from the dimerization and shear-twist motions already presented. 
We simulate the transient powder diffraction patterns for the observable in-plane Miller-indices $\lbrace h0l \rbrace$ in the kinematical approximation using a ZnPc super cell comprising $N=3\times4=12$ unit cells (\textit{c.f.} manuscript figure 2). 
The scattering intensity as a function of the magnitude of the scattering vector $q = \abs{\mathbf{q}}$ for a crystal geometry defined by the atomic positions $\lbrace \mathbf{x}_{s_j} \rbrace$ is given by
\begin{equation}
  \tilde I(\abs{\mathbf{q}}) = \sum_{\lbrace h0l \rbrace} \abs{\sum_j^N \sum_{s_j} f_{s_j}(q) e^{-M_{s_j}} e^{-\i\mathbf{q}\cdot \mathbf{x}_{s_i}} \delta(\abs{\mathbf{q}}-\abs{\mathbf{H}_{h0l}})}^2 .
  \label{eq:S4}
\end{equation}
The double sum runs over all unit cells $j$ and the atoms in that unit cell $s_j$. 
The product of the atomic form factor $f_{s_j}$, the Debye-Waller contribution $e^{-M_{s_j}}$ and  the phase factor $e^{-\i\mathbf{q}\cdot\mathbf{x}_{s_j}}$ gives the contribution from each atom to the total scattered electron wave function. 
Finite size effects of the small crystallites are included by convolving the scattering intensity with a line shape function yielding the powder scattering intensity
\begin{equation}
  I(q) = (\tilde I \ast \Gamma)(q), 
  \label{eq:S5}
\end{equation}
with the Gaussian line shape function 
\begin{equation}
  \Gamma(q) = \frac{e^{-\frac{q^2}{2\sigma^2}}}{2 \pi \sigma},
  \label{eq:S6}
\end{equation}
where $\sigma$ is an empirical parameter determining the line width. 
Considering the excitation density of 1 in 1000 molecules (\textit{c.f.} section \ref{sec:1}), we see that roughly 1 in 83 supercells with $N=12$ will bare a geometric deformation. 
Hence the differential diffracted intensity is given by 
\begin{equation}
  \Delta I(q) = \left(I_\text{X}(q) + 82 \cdot I_\text{G}(q) \right) - 83 \cdot I_\text{G}(q),
  \label{eq:S7}
\end{equation}
where $I_\text{X}$ is the diffraction intensity of a supercell in the excited state, \textit{i.e.} with geometric distortion of the central dimer unit, and $I_\text{G}$ is the diffraction intensity of an undistorted ground state supercell. 
The simulations performed here assume that we have a perfect crystalline powder, while the ZnPc thin films used in these experiments are textured. The crystalline rotational domains share the crystallographic [010]-direction as a common out-of-plane crystal axis. 
This can lead to some deviations between the relative intensities in the simulated and observed changes in diffraction intensity.\par
\begin{figure}[tbp]
  \includegraphics{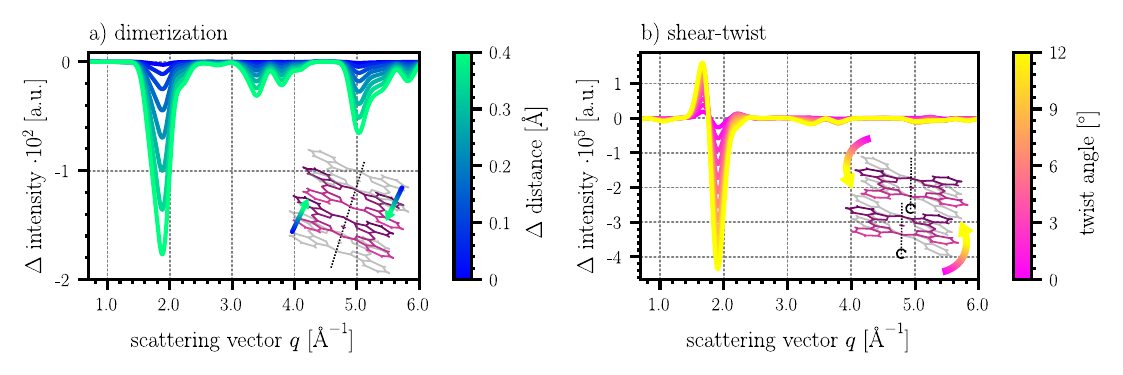}
  \caption{Expected differential diffraction signals for a) dimerization along the crystallographic \textbf{a}-axis and b) a shear-twist motion around a central molecular axis pointing out of the thin film plane.}
  \label{fig:S2}
\end{figure}

Figure \ref{fig:S2} shows the expected differential diffraction signals for a) the dimerization along the crystallographic \textbf{a}-axis as a function of distance decrease between the two monomers and b) the shear-twist motion around a central molecular axis, which is the projection of the crystallographic \textbf{b}-axis onto the molecular plane pointing out of the thin film plane, as a function of twist angle. 
We need to point out that the dimerization signal is much smaller than that of the shear-twist. \par
\begin{figure}[tbp]
  \includegraphics{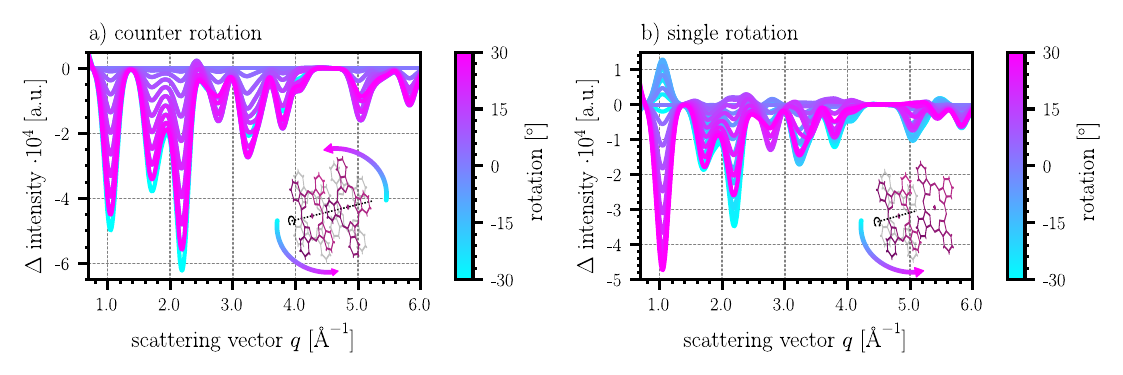}
  \caption{Expected differential diffraction signals for a) counter rotation and b) a single rotation around a normal axis on the molecular plane}
  \label{fig:S3}
\end{figure}
\begin{figure}[tbp]
  \includegraphics{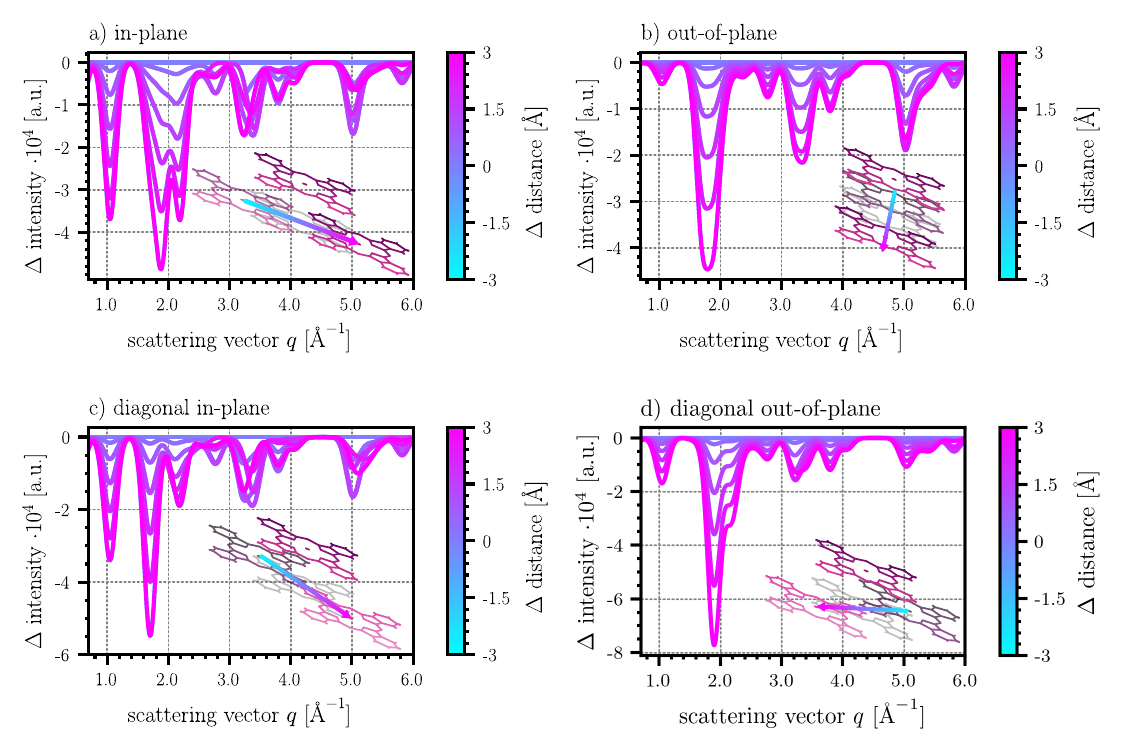}
  \caption{Expected differential diffraction signals for shifting along high symmetry axis of the ZnPc molecule with main contributions in (a) and c)) and out of the thin film plane(b) and d)). Top row (a) and b)) shows movement along two-fold rotation axis in between the isoindole units and bottom row (c) and d)) shows movement along diagonal two-fold rotation axis through the isoindole units.}
  \label{fig:S4}

\end{figure}

While these two movements describe the observed signal very well, many different other geometries could be considered, and indeed several other relaxation pathways along high symmetry path can lead to a potential energy minimum in excimer systems \cite{Casanova_2015,Hoche_2017}. 
In figure \ref{fig:S3} we show the evolution of the differential diffraction signal as a function of twist angle around an axis normal to the molecular plane through the central Zn atom. 
In a) a counter rotation of the two central dimer molecules is considered, while in b) only one of the molecules rotates and the other one is kept fix. 
Figure \ref{fig:S4} shows the signal for several shifts along axes within the molecular plane. 
The subfigures a) and b) show movements along a the two high symmetry axes cutting the phthalocyanine in half in between the isoindole units. 
One axis lies within the thin film plane (a) while the other one  points out of it (b). 
Other important high symmetry axes are the two diagonal two-fold rotation axis and the respective signal change for movement mostly within the thin film plane is shown in c) while the signal change for the movement along the axis mainly pointing out of the thin film plane is shown in d). \par
For all these movements, the expected signal does not reproduce the observed signal increase and decrease signature around $q \approx \qty{1.9}{\per\angstrom}$. 
Also a strong signal decrease is found at $q = \qty{1.0}{\per\angstrom}$, which is associated with a signal decrease in the (002) diffraction intensity. 
This feature is not observed in the data. 
Hence we conclude that these geometries are not contributing to the overall geometric relaxation.

\FloatBarrier
\section{ZnPc: Time trace fitting}
\begin{figure}[tbp]
  \includegraphics{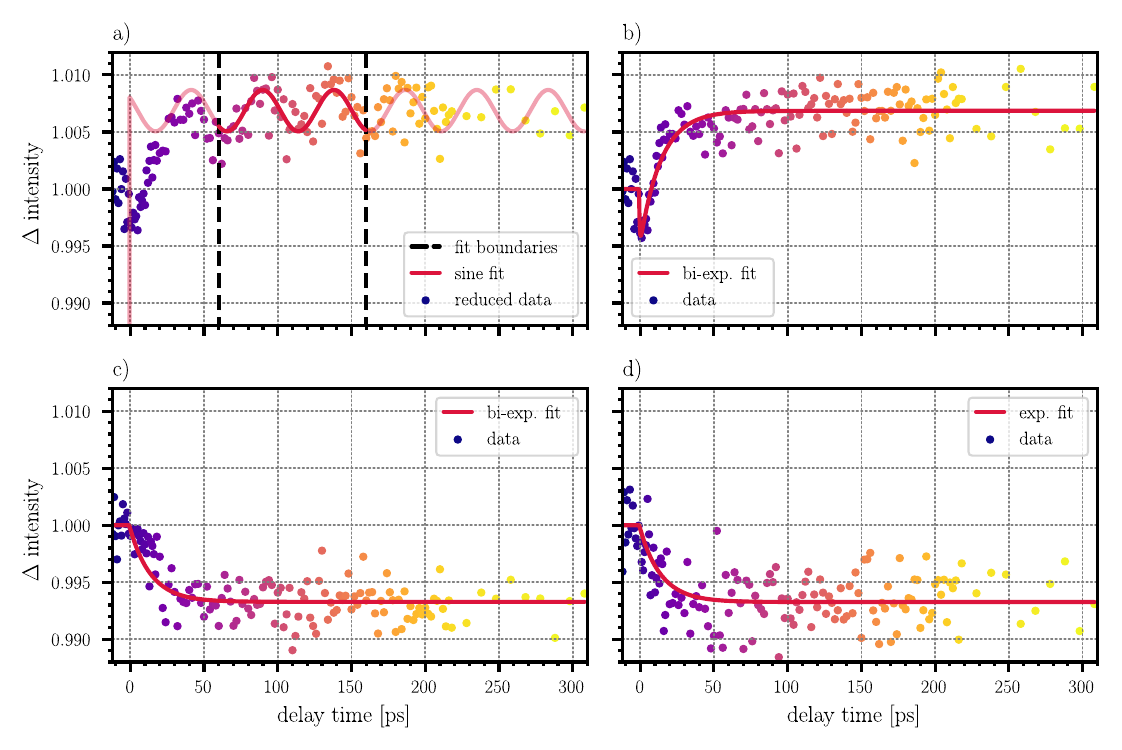}
  \caption{Time trace fitting of ZnPc differential diffraction intensity. a) Fitting of the coherent acoustic phonon ringing. b) to d) Fit of kinetic model to features P1, P2 and P3, respectively.}
  \label{fig:S5}
\end{figure}
As described in the main manuscript the three features P1, P2 and P3 in the differential diffraction signal have been analyzed by a kinetic model of exponential functions with two time constants. 
To fit an bi-exponential function to P1 the data needs to be corrected for the coherent ringing. 
For that purpose we fitted a sine function 
\begin{equation}
  f_0(t) = A \sin \left(\omega (t-t_0) + \Phi \right) + I_0
  \label{eq:S8}
\end{equation}
in the range $\qty{60}{\ps} \leq t \leq \qty{160}{\ps}$ where the ringing is most prominent (\textit{c.f.} figure \ref{fig:S5} a)). 
We find an angular frequency of $\omega = \qty{130(7)}{\giga\hertz}$ and a phase shift of $\Phi = \qty{1.2(3)}{\pi}$. 
Under the assumption that the lowest order out-of-plane acoustic phonon is the cause for the ringing, the speed of sound can be extracted from the oscillation frequency as shown in\cite{Chatelain_2014}. We approximate the film as a free-standing layer, hence open boundary conditions apply, which determines to wavelength to $\lambda = 2 d$, where $d$ is the film thickness. This yields
\begin{equation}
  v_s = \lambda f =2d f = \frac{d \omega}{\pi} = \qty{2.07(11)}{\km\per\second}
  \label{eq:S9}
\end{equation}
as the sound velocity with $d=\qty{50}{\nm}$. 
We extend the sine function \eqref{eq:S8} over the data range $t \geq t_0$ (\textit{c.f.} figure \ref{fig:S5} a)) and subtract it from the data with $I_0 = 0$. 
The resulting data is shown in figure \ref{fig:S5} b) and is used for fitting the kinetic model. 
We fit the following kinetic model to the time traces of P1, P2 and P3:
\begin{align}
  I_\text{P1}(t) &= 
  \begin{cases}
  A_{11}\left( \exp\left( - \frac{t-t_0}{\tau_1}\right) -1 \right) + A_{12}\left( \exp\left( - \frac{t-t_0}{\tau_2}\right) -1 \right) + 1  & t > t_0 \\
  1 & t \leq t_0   
  \end{cases} \\
  I_\text{P2}(t) &= 
\begin{cases}
A_{21}\left( \exp\left( - \frac{t-t_0}{\tau_1}\right) -1 \right) + A_{22}\left( \exp\left( - \frac{t-t_0}{\tau_2}\right) -1 \right) + 1  & t > t_0 \\
1 & t \leq t_0   
\end{cases} \\
I_\text{P3}(t) &=    
\begin{cases}
A_3\left( \exp\left( - \frac{t-t_0}{\tau_1}\right) -1 \right) + 1  & t > t_0 \\
1 & t \leq t_0   
\end{cases}
\end{align}
We us a single residual function and global parameters for the time constants $\tau_1$ and $\tau_2$. 
The temporal offset was determined to $t_0 = \qty{-0.7713}{\ps}$ and was kept fix for all fitting procedures.
The fits were performed using the \texttt{lmfit python} package \cite{lmfit_2023}.
Figures \ref{fig:S5} b) to d) show the resulting fitting curves.
\FloatBarrier
\section{Potential Energy Landscape from Density Functional Theory}
The self-consistent field (SCF) calculations on the isolated dimer were carried out using the \texttt{QuantumESPRESSO} suite. 
To maximize computational efficiency, ultrasoft norm-conserving pseudopotentials were used to map the ground and excited state energy surfaces \cite{Hamann2013}, where we again employed the Grimme-D3 van der Waals dispersion correction with BJ damping as in Section S2. 
To determine the excited state, we employed the method of Marini \emph{et al} \cite{Marini_2021}, placing different chemical potentials for the holes in the valence bands and excited electrons in the conduction bands.
This required Marzari-Vanderbilt-DeVita-Payne cold smearing \cite{Marzari_1999} with a Gaussian spread of \qty{1.8374}{\milli\rydberg} (equals \qty{290}{\kelvin}) for both valance and conduction band manifold. 
These electronic structure calculations were performed at the Gamma point with a kinetic and wave function cutoff energy of \qty{660}{\rydberg} and \qty{90}{\rydberg}, respectively.

First we performed a relaxation of the monomer geometry in the ground and excited state with one electron in the conduction band manifold. 
For that purpose, a ZnPc monomer was placed in the center of a cubic unit cell with a lattice constant of \qty{45}{\angstrom}, roughly three times the molecular diameter, to isolate the molecule within the periodic boundaries. 
We used the Martyna-Tuckermann correction \cite{Martyna_1999} to correct for long range forces in an isolated cluster. We found no significant difference between the excited and ground state intra-molecular geometry for the monomer. These parameters were then used in all subsequent calculations.

In a next step, we tuned the intra-molecular geometry of each monomer in an excited dimer setting to encourage delocalization of the wave function over both monomers, a key feature of an excimeric state.
It has been shown, that mixing the ground $\vec{R}(S_0)$ and excited state geometry $\vec{R}(S_1)$ symmetrically in the comprising monomers is a key requirement to reproduce absorption spectra in molecular clusters \cite{Craciunescu_2022,Craciunescu_2023}.
We scanned the geometry mixing ratio  $(1-\chi)\vec{R}(S_0) + \chi \vec{R}(S_1)$ in excited state SCF calculations with one electron in the conduction band manifold and found the energetic minimum for $\chi = 0.5$, as expected for an evenly spread excitation over a dimer.
We used the resulting inter-molecular geometry for all following calculations. \par
We then proceeded to systematically explore the ground and excited state potential energy landscape performing a ground or excited state SCF calculation for each dimer geometry. 
We reduced the inter-molecular distance by $\Delta d_\text{Zn-Zn}$ from $\Delta d_\text{Zn-Zn} = \qty{0}{\angstrom}$ to $\Delta d_\text{Zn-Zn} = \qty{0.4}{\angstrom}$ in steps of \qty{0.025}{\angstrom} and varied the twist angle from \qty{0}{\degree} to \qty{17}{\degree} in steps of \qty{1}{\degree} yielding us a point grid of $17 \times 18 = 306$ points for ground and excited state, respectively. 

\FloatBarrier
\section{Fluorinated ZnPc: Diffraction, transient diffraction and time traces}
\begin{figure}[tbp]
  \includegraphics{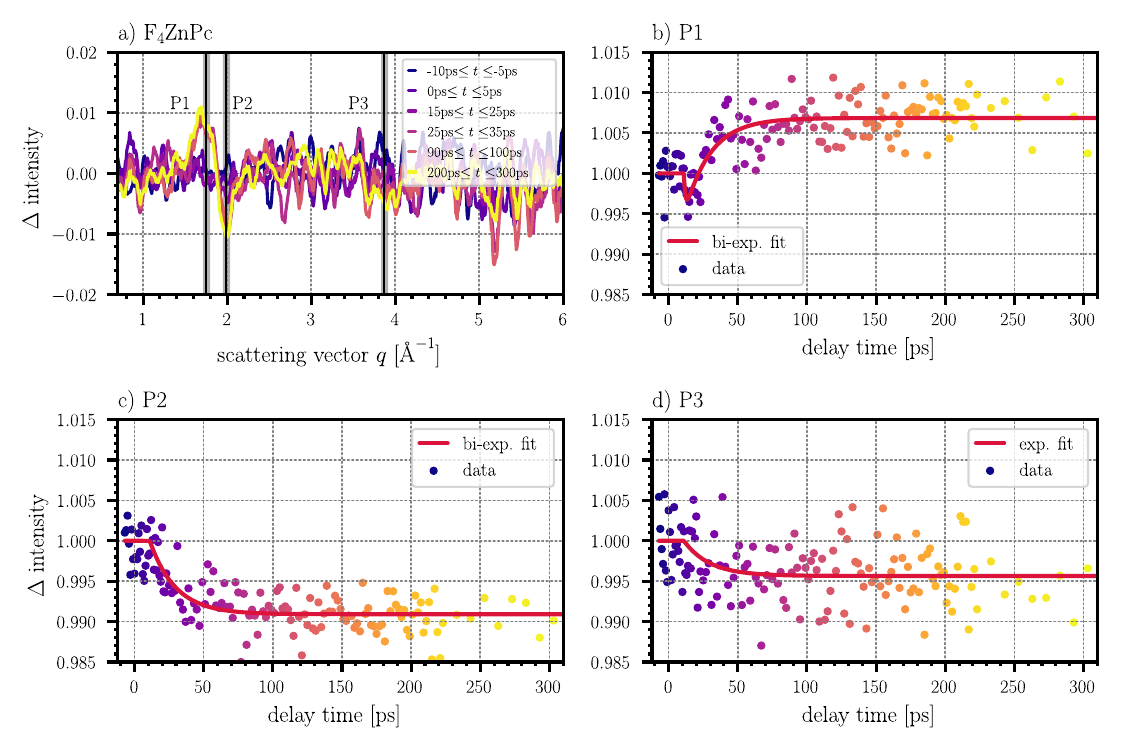}
  \caption{F$_4$ZnPc: Transient diffraction signal a) and time traces of features P1, P2 and P3 in b), c) and d), respectively}
  \label{fig:S6}
\end{figure}

\begin{figure}[tbp]
  \includegraphics{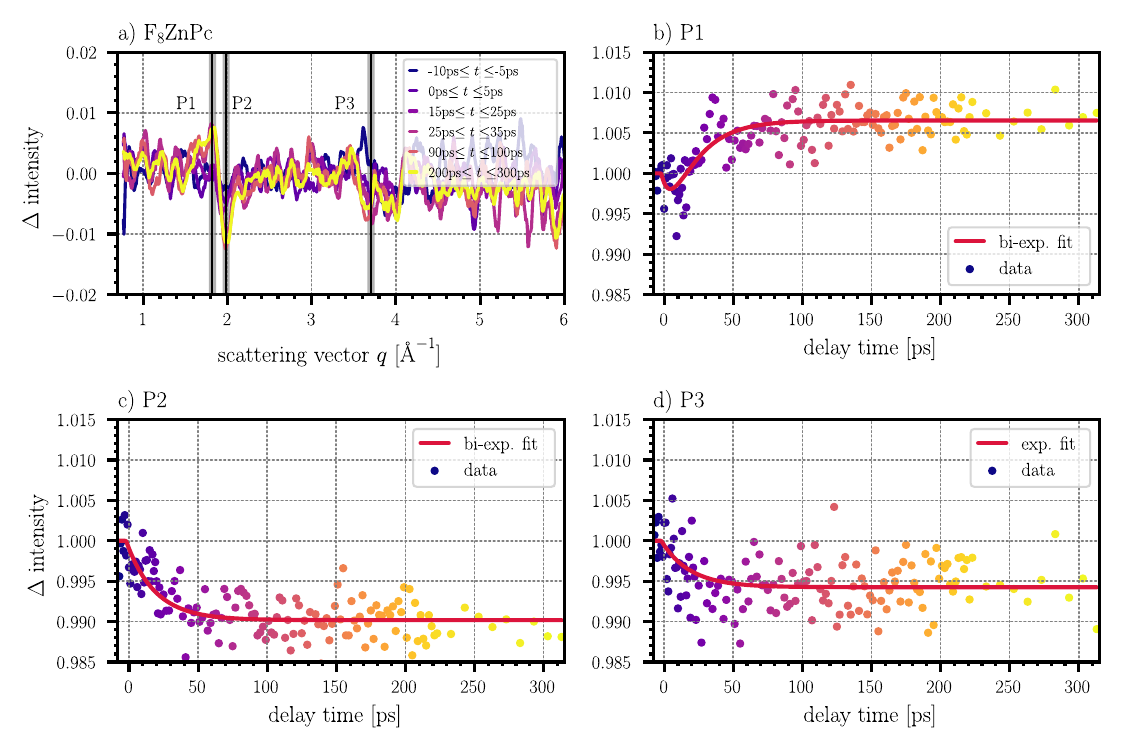}
  \caption{F$_8$ZnPc: Transient diffraction signal a) and time traces of features P1, P2 and P3 in b), c) and d), respectively}
  \label{fig:S7}
\end{figure}

Figures \ref{fig:S6} and \ref{fig:S7} show the transient diffraction data for a \qty{50}{\nm} F$_4$ZnPc and F$_8$ZnPc thin film, respectively. 
We find the same behavior in the transient diffraction of the fluorinated films we observe in ZnPc. 
The time traces through the ZnPc equivalent features P1 to P3 are shown in the respective subfigures b) to d). 
The same fitting routine explained above in section 4 has been applied to the data and the time constants are plotted in figure 5 of the main manuscript.

\FloatBarrier
\printbibliography